\documentclass[journal=jctcce,manuscript=article]{achemso}
\setkeys{acs}{articletitle=true}
\usepackage[utf8]{inputenc}
\usepackage{geometry}
\usepackage{chemfig}

\usepackage{amsmath}
\usepackage{amsfonts}
\usepackage{amssymb}
\usepackage{amsthm}
\usepackage{mathtools} 
\usepackage{bm}        
\usepackage{upgreek}   

\usepackage{graphicx}
\usepackage[center]{subfigure} 
\usepackage{adjustbox}
\usepackage{booktabs}
\usepackage{multirow}
\usepackage{xcolor}

\usepackage{achemso}   
\usepackage{comment}   
\usepackage{xparse}    

\newlength\tindent
\newcommand*{\diff}{\mathop{}\!\mathrm{d}}
\newcommand{\abs}[1]{\left| {#1} \right|}

\newcommand{\tr}[1]{\text{Tr }{#1}}

\newcommand\cx{c\ped{x}}

\newcommand\epsilonxc{\varepsilon\ped{xc}}
\newcommand\epsilonx{\varepsilon\ped{x}}
\newcommand\epsilonc{\varepsilon\ped{c}}

\NewDocumentCommand{\bh}{G{}G{}}{\tr{\mathbf{h}^{#1}\mathbf{D}^{#2}}}
\NewDocumentCommand{\bV}{G{}G{}}{\tr{\mathbf{V}^{#1}\mathbf{D}^{#2}}}
\NewDocumentCommand{\bJ}{G{}G{}}{\tr{\mathbf{D}^{#1}\mathbf{J}(\mathbf{D}^{#2})}}
\NewDocumentCommand{\bK}{G{}G{}}{\tr{\mathbf{D}^{#1}\mathbf{K}(\mathbf{D}^{#2})}}
\NewDocumentCommand{\bKLR}{G{}G{}}{\tr{\mathbf{D}^{#1}\mathbf{K}^{\text{LR}}(\mathbf{D}^{#2})}}

\NewDocumentCommand{\bex}{G{}G{}}{\int \rho^{#1}(\mathbf{r})\epsilonx(\rho^{#2}(\mathbf{r}))\diff \mathbf{r}}
\NewDocumentCommand{\bec}{G{}G{}}{\int \rho^{#1}(\mathbf{r})\epsilonc(\rho^{#2}(\mathbf{r}))\diff \mathbf{r}}
\NewDocumentCommand{\bexc}{G{}G{}}{\int \rho^{#1}(\mathbf{r})\epsilonxc(\rho^{#2}(\mathbf{r}))\diff \mathbf{r}}
\NewDocumentCommand{\bvxc}{G{}G{}}{\int v\ped{xc}(\rho^{#1}(\mathbf{r}))\chi_\mu(\mathbf{r})\chi_\nu(\mathbf{r})\diff \mathbf{r}}
\NewDocumentCommand{\bvx}{G{}G{}}{\int v\ped{x}(\rho^{#1}(\mathbf{r}))\chi_\mu(\mathbf{r})\chi_\nu(\mathbf{r})\diff \mathbf{r}}
\NewDocumentCommand{\bvc}{G{}G{}}{\int v\ped{c}(\rho^{#1}(\mathbf{r}))\chi_\mu(\mathbf{r})\chi_\nu(\mathbf{r})\diff \mathbf{r}}

\newcommand\ap[1]{\ensuremath{^{\text{#1}}}}
\newcommand\ped[1]{\ensuremath{_{\text{#1}}}}

\graphicspath{{images/}} 

\author{Alberto Barlini}
\affiliation{Scuola Normale Superiore,
             Piazza dei Cavalieri 7, 56126 Pisa, Italy.}
\author{Julien Bloino}
\affiliation{Scuola Normale Superiore,
             Piazza dei Cavalieri 7, 56126 Pisa, Italy.}
\author{Henrik Koch}
\affiliation{Department of Chemistry, Norwegian University of Science and Technology, Trondheim, Norway}             
\author{Tommaso Giovannini}
\affiliation{Department of Physics, University of Rome Tor Vergata, and INFN, Via della Ricerca Scientifica 1, 00133, Rome, Italy }
\email{tommaso.giovannini@uniroma2.it}

\title[]{Multilevel DFT Response Theory}

\begin{document}

\begin{abstract}
    We present a general computational protocol for the evaluation of extensive molecular response properties in complex environments within a polarizable quantum embedding framework. The approach extends multilevel density functional theory (MLDFT) to response theory by formulating the coupled-perturbed Kohn–Sham (CPKS) equations for the MLDFT Hamiltonian. The method is further coupled to an additional polarizable molecular mechanics layer based on the fluctuating-charge (FQ) force field, which allows an accurate yet computationally efficient description of long-range interactions. We apply this new protocol to compute static and frequency-dependent linear polarizabilities and first hyperpolarizabilities of para-nitroaniline (PNA) in 1,4-dioxane and 3-hydroxybenzoic acid (HBA) in aqueous solution. The framework enables physicochemical insight into solute-solvent interactions by disentangling the competing roles of electrostatics, mutual polarization, and quantum confinement (Pauli repulsion). The results match available experiments, demonstrating the reliability and robustness of the proposed approach and providing a viable route for response properties within quantum embedding methods.
\end{abstract}

\begin{tocentry}
\includegraphics[]{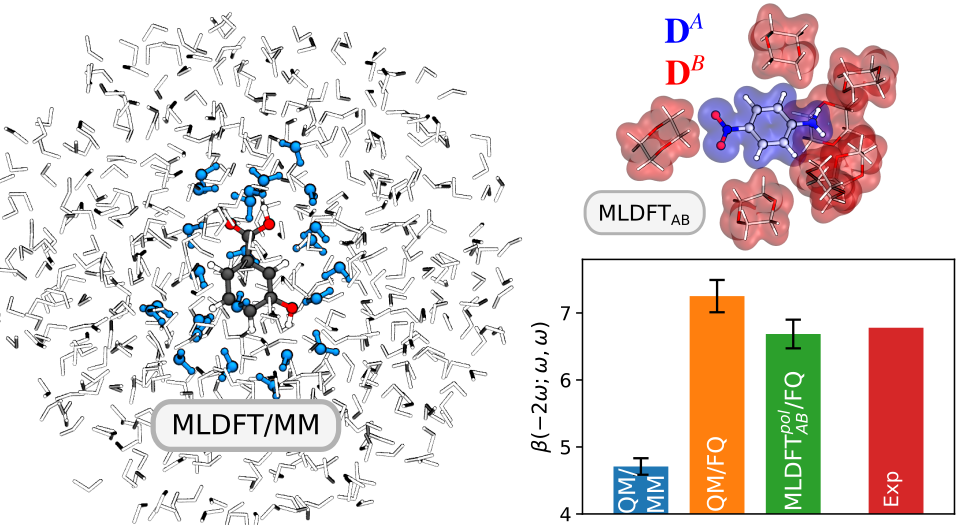}
\end{tocentry}

\section{Introduction}

Molecular response properties are key quantities underlying a broad range of applications, ranging from photonics to molecular sensing and energy-related technologies.\cite{gu2016molecular,castet2013design,radhakrishnan2008molecular,marks1995design,papadopoulos2006non,boyd2003nonlinear} When molecular systems are in the condensed phase, their response properties can be substantially altered by the surrounding medium.\cite{tomasi2005,tomasi2002molecular,quertinmont2015explicit,dellai2025solvent,cammi2013molecular} In this context, theoretical approaches play a pivotal role in complementing and rationalizing experimental measurements,\cite{cammi2000attempt,castet2022predicting} as well as in enabling the in-silico design of functionalized molecules and materials.\cite{kanis1994design,marks1995design} However, the accurate modeling of response properties in complex environments still remains a major challenge in computational chemistry.\cite{mennucci2019multiscale} Indeed, fully quantum mechanical (QM) treatments rapidly become computationally prohibitive as system size grows, due to the large number of degrees of freedom that must be accounted for. This limitation has motivated the development of the so-called focused models.\cite{warshel1972calculation,warshel1976theoretical,tomasi2005,giovannini2020csr,mennucci2019multiscale,giovannini2023continuum} In these hybrid schemes, the total system is partitioned into regions described at different levels of theory,\cite{giovannini2020csr,mennucci2019multiscale,tomasi2005,warshel1972calculation,warshel1976theoretical,senn2009qm,lin2007qm} restricting the QM description to a selected subsystem (active region) while the surrounding environment is treated at a lower level.

Among the focused models, to achieve a fully atomistic description of the total system, Quantum Mechanics/Molecular Mechanics (QM/MM) approaches can be exploited, providing an effective balance between accuracy and computational cost.\cite{warshel1972calculation,warshel1976theoretical,senn2009qm,lin2007qm} Here, the environment is described through classical electrostatic force fields, which can be refined by accounting for mutual polarization effects in polarizable embedding. However, except for a limited number of approaches\cite{giovannini2017disrep,giovannini2019quantum,reinholdt2017polarizable,curutchet2018density}, purely quantum interactions, such as Pauli repulsion and dispersion, are usually introduced through parameterized functions and therefore do not affect molecular response properties. Such a limitation can lead to an unphysical description of response properties, particularly because Pauli repulsion may effectively act as a quantum confinement on the target electronic density.\cite{giovannini2019quantum,marrazzini2020calculation}

To address these shortcomings, quantum embedding strategies can be exploited, in which both the target system and its surroundings are described at the QM level. \cite{gordon2013accurate,gordon2007effective,sun2016quantum,knizia2013density,chulhai2018projection,chulhai2017improved,wen2019absolutely,ding2017embedded,goodpaster2012density,manby2012simple,goodpaster2010exact,zhang2020multi,ramos2015performance,pavanello2011modelling,saether2017density,hoyvik2020convergence,myhre2014multi,myhre2016multilevel} Within this class of methods, multilevel density functional theory (MLDFT) \cite{marrazzini2021multilevel,giovannini2023integrated,giovannini2024tdmldft} is a promising quantum embedding approach that effectively partitions the system into active and inactive QM fragments by employing a density matrix decomposition scheme based on Cholesky decomposition.\cite{aquilante2011cholesky,sanchez2010cholesky,koch2003reduced,christiansen2006coupled,saether2017density,giovannini2021mlhf_ab,goletto2022linear} Such a strategy overcomes challenges associated with non-additive kinetic energy terms encountered, for instance, in Frozen Density Embedding (FDE) methods \cite{goodpaster2012density,tamukong2014density,tamukong2017accurate,wesolowski2004hydrogen,bennie2017pushing,lee2019projection,neugebauer2005merits,wesolowski2015frozen} and provides significant computational savings by restricting calculations to active molecular orbitals (MOs). As in other embedding frameworks,\cite{egidi2021polarizable,lafiosca2024multiscale} MLDFT can be further combined with a third MM layer to capture long-range electrostatic effects in an accurate yet cost-effective manner.\cite{giovannini2023integrated,giovannini2024tdmldft} In particular, when coupled to a polarizable MM embedding based on the fluctuating charge (FQ) force field,\cite{giovannini2020csr,rick1994dynamical} MLDFT has been shown to provide an accurate description of excitation energies\cite{giovannini2024tdmldft} and ground-state properties of radicals in solution.\cite{giovannini2023integrated}

In this work, we propose a general computational protocol for evaluating response properties within a polarizable MLDFT/MM framework. While the method is general and any polarizable MM force field can be exploited, in this work, we consider the coupling of MLDFT with the polarizable FQ force field. The approach is formulated in terms of coupled-perturbed Kohn-Sham (CPKS) equations, thus providing a general route to field-derivative properties in embedded systems. Here, we define it for extensive response properties, such as electric polarizability and first hyperpolarizability, for which a proper localization of the active MOs within the selected active region is essential to obtain a physically meaningful partitioning of the response.\cite{goletto2022linear} Such localization is achieved by exploiting the recently introduced Kohn–Sham Fragment-Localized MOs (KS-FLMOs) procedure.\cite{giovannini2024kohn} KS-FLMOs are obtained via an energy-based minimization of fragment electronic energies and correspond to the MOs that are maximally localized within the spatial domain of each fragment. Therefore, by combining KS-FLMOs with CPKS-MLDFT formulation, we introduce a polarizable MLDFT/FQ framework that properly accounts for mutual active–inactive polarization, as well as quantum confinement effects.

The paper is organized as follows. First, we describe the theory to predict the response properties of embedded systems within the MLDFT/FQ framework. Next, we provide the computational details and then present numerical applications to para-nitroaniline (PNA) in 1,4-dioxane and to 3-hydroxybenzoic acid (HBA) in aqueous solution. Finally, the manuscript concludes with a summary of the main results, along with conclusions and future perspectives.

\section{Theoretical Approach} \label{sec:theory}

\begin{figure}[!htbp]
    \centering
    \includegraphics[width=.4\linewidth]{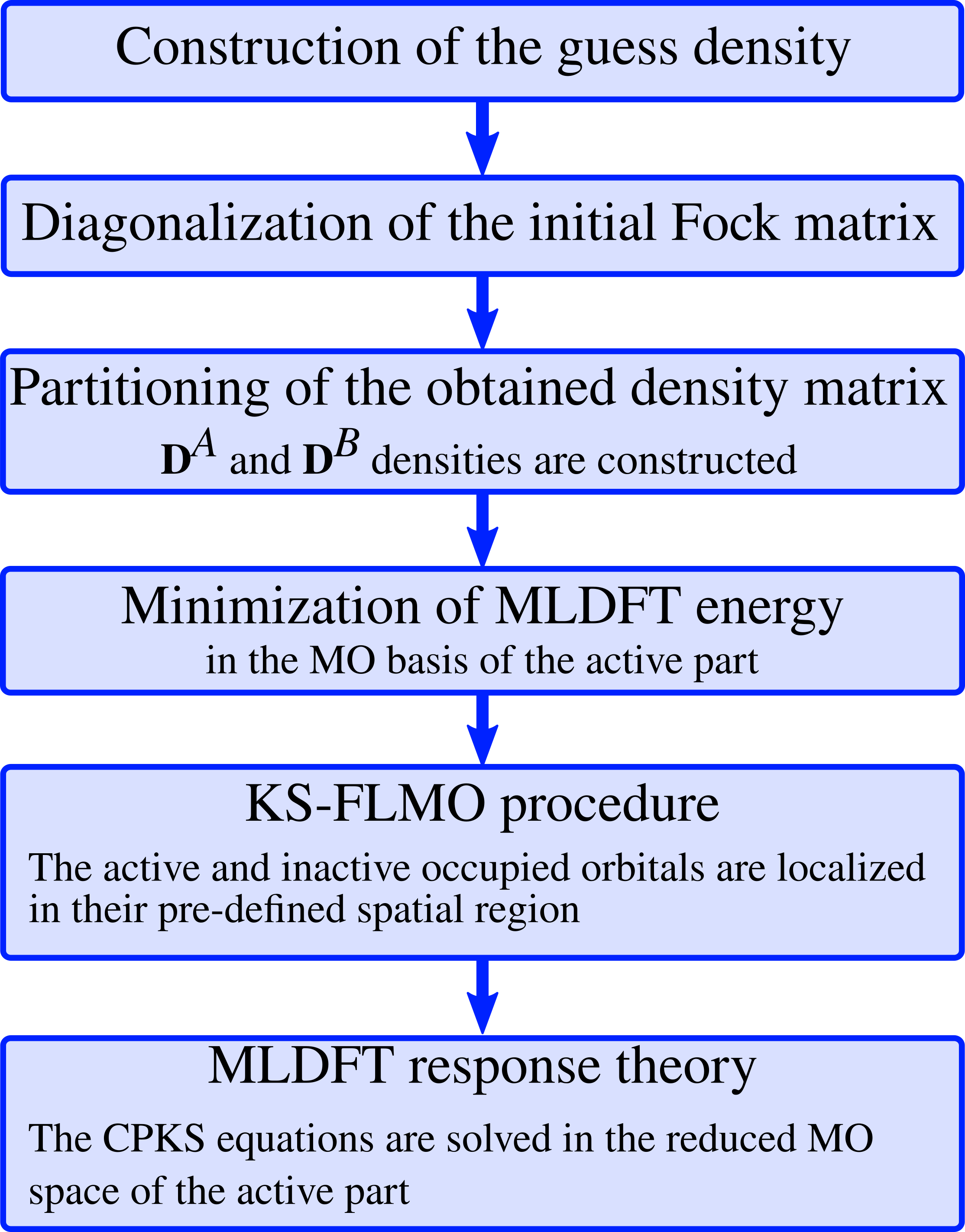}
    \caption{Graphical view of the computational procedure.}
    \label{fig:comput_procedure}
\end{figure}

The calculation of response properties within a polarizable MLDFT/MM framework involves several computational steps, which are summarized in Fig. \ref{fig:comput_procedure}. These steps are general and can be applied to any property of a molecular system embedded in an external environment described at the MLDFT/MM level. We briefly remark that, in the three-layer MLDFT/MM scheme, the overall system is partitioned into a quantum region treated at the MLDFT level and a classical part which is described fully atomistically by means of (polarizable) MM force fields. The quantum region is, in turn, partitioned into active (A) and inactive (B) subsystems, which generally represent the target system and the closest, strongly interacting molecules of the embedding. In the following, we discuss these steps with a specific focus on the calculation of linear and non-linear response properties to an oscillating electric field.

\paragraph{Construction of the guess density} The quantum region described at the MLDFT level is partitioned into $N_f$ fragments. These fragments typically correspond to the molecules forming the active and inactive MLDFT subsystems (i.e., the target molecule and, in solution, the surrounding solvent molecules). To obtain a good starting guess, for each fragment, the DFT energy is minimized within the fragment atomic orbital (AO) basis, and the density matrix $\mathbf{D}$ of the entire quantum (MLDFT) region is then constructed as the direct sum of the molecular densities of the fragments:\cite{giovannini2024tdmldft,goletto2022linear}
\begin{equation}
\mathbf{D} = \bigoplus^{N_f}_i \mathbf{D}_i
\label{eq:dens_direct_sum}
\end{equation}

Thus, the resulting density matrix is block-diagonal, with blocks given by the self-consistent field (SCF)-converged fragment density matrices $\mathbf{D}_i$.

\paragraph{Diagonalization of the initial Fock matrix} Using the density matrix defined in Eq. \ref{eq:dens_direct_sum}, we construct the initial Fock matrix of the total quantum region and diagonalize it to obtain an idempotent, physically consistent density matrix.\cite{saether2017density} At this stage, a portion of the active–inactive polarization energy is already introduced. Indeed, in previous works, we have shown that, for neutral species, the largest part of the active–inactive polarization energy is captured at this step. \cite{marrazzini2021multilevel,giovannini2023integrated}

\paragraph{Partitioning of the obtained density matrix} Following the diagonalization of the initial Fock matrix, a new density matrix for the total system is obtained, which serves as the starting point of the MLDFT procedure. The density matrix is decomposed into active and inactive parts, $\mathbf{D} = \mathbf{D}^A + \mathbf{D}^B$, by performing a partial Cholesky decomposition of the occupied space associated with the active region.\cite{aquilante2011cholesky,sanchez2010cholesky,koch2003reduced,christiansen2006coupled,giovannini2021mlhf_ab,goletto2022linear} The Cholesky decomposition of the total density $\mathbf{D}$ into $\mathbf{D}^A$ and $\mathbf{D}^B$ is unique provided that the same set of pivots is employed. In multilevel schemes, the pivots are chosen among the diagonal elements corresponding to basis functions centered on the active atoms.\cite{saether2017density,sanchez2010cholesky,myhre2014multi} Note that, in principle, the partitioning of the density matrix is not unique, and other strategies can be exploited. However, by using the aforementioned procedure based on a partial Cholesky decomposition, the MOs associated with active and inactive subsystems remain orthogonal throughout all subsequent SCF iterations.~\cite{sanchez2010cholesky}
As a result of the decomposition, the active Cholesky MOs are obtained and used to construct the active density matrix, $\mathbf{D}^A$. The active virtual orbitals are instead generated in terms of projected AOs (PAOs), built from the AOs centered on the active region and orthonormalized via the L\"owdin procedure. \cite{hoyvik2015perspective,saebo1993} A threshold needs to be employed to remove linear dependencies, which was set as default to $10^{-6}$, sufficient for the cases considered in this work. Finally, the inactive density matrix, $\mathbf{D}^B$, is obtained as the difference between the total density $\mathbf{D}$ and the active contribution $\mathbf{D}^A$.

\paragraph{Minimization of the MLDFT/MM Energy} 

The MLDFT/MM energy reads: \cite{giovannini2023integrated,giovannini2024tdmldft}
\begin{equation}
    \mathcal{E} = E\ped{MLDFT} + E\ped{MM} + E\ap{int}\ped{MLDFT/MM}
    \label{eq:energy_mldft_mm}
\end{equation}
where $E\ped{MLDFT}$ denotes the MLDFT energy, while $E\ped{MM}$ and $E\ap{int}\ped{MLDFT/MM}$ are the MM energy and the MLDFT/MM interaction energy, respectively.

The MLDFT energy $E\ped{MLDFT}$ can be expressed as:
\begin{equation} \label{eq:dft_energy_total}
    E\ped{MLDFT}[\mathbf{D}^A; \mathbf{D}^B] = E^{A}[\mathbf{D}^A] + E^{B}[\mathbf{D}^B] + E^{AB}[\mathbf{D}^A; \mathbf{D}^B] 
\end{equation}

where $E^{X}[\mathbf{D}^{X}]$ is the energy of the subsystem $X = A, B$ and $E^{AB}[\mathbf{D}^A; \mathbf{D}^B]$ is the active-inactive interaction energy. The contribution $E^{X}[\mathbf{D}^{X}]$ can be written as follows: \cite{giovannini2024tdmldft}
\begin{equation} \label{eq:dft_energy_x}
    \begin{split}
        E^X[\mathbf{D}^{X}] &= \bh{X}{X} + \dfrac{1}{2}\bJ{X}{X} \\
                          & - \dfrac{1}{2}\cx\bK{X}{X} \\
                          &+ (1-\cx)\bex{X}{X}  \\
                          &+ \bec{X}{X} \\
                          & - \frac{c_{\text{LR}}}{2} \tr{\mathbf{D}\mathbf{K}_{\text{LR}}(\mathbf{D})}
    \end{split}
\end{equation}
where $\mathbf{J}$ and $\mathbf{K}$ are the Coulomb and exchange matrices, respectively, and $\mathbf{h}^X=\mathbf{T}+\mathbf{V}^X$ is the one-electron term, accounting for the kinetic energy $\mathbf{T}$ and the electron–nuclear attraction within fragment $X$ ($\mathbf{V}^X$). The corresponding electron density $\rho^X(\mathbf{r})$ is obtained from the density matrix $\mathbf{D}^X$ in the usual AO representation. Notably, the density matrix partitioning enables a direct decomposition of the kinetic energy contribution, thereby avoiding the difficulties associated with non-additive kinetic energy terms in FDE methods.\cite{wesolowski2022non,wesolowski1996accuracy,de2012exact} $\epsilon_x$ and $\epsilon_c$ are the exchange and correlation energy densities per particle. The coefficient $c_x$ determines whether a pure DFT functional ($c_x=0$) or a hybrid functional ($c_x\neq 0$) is employed. $\mathbf{K}_{\text{LR}}$ denotes the long-range Hartree-Fock (HF) exchange matrix used in range-separated functionals, written in terms of the $\mathrm{erf}(\gamma r_{ij})/r_{ij}$ operator, where $\gamma$ and $c_{\text{LR}}$ define the chosen range-separated functional.\cite{chai2008long,yanai2004new,rohrdanz2009long}

The active-inactive interaction energy can be written as:\cite{giovannini2024kohn}
\begin{equation} \label{eq:dft_energy_ab}
    \begin{split}        
        E^{AB}[\mathbf{D}^A; \mathbf{D}^B] & = \bV{A}{B} + \bV{B}{A} \\
                                           & + \bJ{A}{B} - \cx \bK{A}{B}  \\
                                           & + \bexc{A}{B} \\
                                           & + \bexc{B}{A} \\ 
                                           & - c_{\text{LR}} \ \bKLR{A}{B} + E^{AB}_{\text{non-add}}  
    \end{split}
\end{equation}
where $\varepsilon_{xc}$ indicates the exchange-correlation energy functional including the $\cx$ parameter. The last term $E^{AB}_{\text{non-add}}$ arises from the non-linearity of $\epsilonx$ and $\epsilonc$ (see Refs. \citenum{marrazzini2020calculation} and \citenum{giovannini2024tdmldft} for its definition).
The computational advantage of MLDFT arises from keeping the density matrix of the inactive subsystem, $\mathbf{D}^B$, frozen during the SCF procedure. As a consequence, the inactive fragment B acts as an external field on the active subsystem A. Thus, the MLDFT Fock matrix is obtained by functional differentiation with respect to the active density only (see Refs.~\citenum{marrazzini2021multilevel} and \citenum{giovannini2024tdmldft} for its explicit definition). A key advantage over full DFT is that the SCF procedure is carried out solely in the MO basis of the active subsystem.\cite{saether2017density} The resulting reduction in the dimensionality of the problem translates into a lower computational cost, which is particularly advantageous for the evaluation of response properties considered in this work.

The integration of a third MM layer can be achieved within Eq.~\ref{eq:energy_mldft_mm} by modeling the MLDFT/MM coupling either as a purely electrostatic embedding, using fixed charges assigned to each MM atom, or through a polarizable embedding, which provides a more physical description of target–environment interactions.\cite{senn2009qm} Among the various polarizable embedding schemes, \cite{olsen2010excited,olsen2011molecular,olsen2015polarizable,reinholdt2017polarizable} as stated above, we adopt the polarizable FQ force field. In this model, each MM atom carries a charge that adapts to the external electrostatic potential and to differences in atomic electronegativities. \cite{rick1994dynamical,rick1995fluctuating,rick1996dynamical} The FQs are obtained by minimizing the following energy functional:\cite{gomez2023multiple}
\begin{equation} \label{eq:QMFQ}
    \begin{split}
        \mathcal{E}[\mathbf{D}_A,\mathbf{D}_B,\mathbf{q},\boldsymbol{\mathbf{\lambda}}] & =  E\ped{MLDFT}[\mathbf{D}_A,\mathbf{D}_B]    \\
        & + \frac{1}{2}\mathbf{q}_\lambda^\dagger\mathbf{M}\mathbf{q}_\lambda + \mathbf{q}_\lambda^\dagger\mathbf{C}_Q  \\
        & + \mathbf{q}_\lambda^\dagger\mathbf{V}(\mathbf{D}) 
    \end{split}
\end{equation}
where $\mathbf{q}_\lambda$ collects the FQ charges together with the Lagrange multipliers enforcing charge conservation on each fragment of the FQ layer. The matrix $\mathbf{M}$ describes the interactions among the FQ charges and includes the blocks associated with the Lagrange multipliers.\cite{lipparini2012linear} The vector $\mathbf{C}_Q$ accounts for electronegativity terms and the charge constraints defined for each FQ moiety. Finally, the term $\mathbf{q}_\lambda^\dagger \mathbf{V}(\mathbf{D})$ represents the MLDFT–FQ interaction, expressed through the electrostatic potential generated by the total density matrix (active + inactive) acting on the charges. The FQ charges are then equilibrated by solving the following set of linear equations:\cite{lipparini2012linear}
\begin{equation} \label{eq:linearqmfq}
    \mathbf{M}\mathbf{q}_{\lambda} = -\mathbf{C}_Q - \mathbf{V}(\mathbf{D}) 
\end{equation}
The MLDFT/MM Fock matrix is finally defined as:
\begin{equation} \label{eq:fock-qmmm}
    F_{\mu\nu}\ap{MLDFT/MM} = F\ap{MLDFT}_{\mu\nu} + \sum_i q_i V_{i,\mu\nu}  
\end{equation}
where $F\ap{MLDFT}_{\mu\nu}$ is the MLDFT Fock. Since the FQs depend on the density \textit{via} Eq. \ref{eq:linearqmfq}, the MLDFT/FQ contribution must be updated at each SCF cycle. This represents the mutual polarization interactions between MLDFT and FQ layers.~\cite{giovannini2023integrated}

\paragraph{Kohn-Sham Fragment-Localized MO Procedure}

As in Multilevel Hartree-Fock (MLHF),\cite{giovannini2021energy,giovannini2022fragment} the occupied MOs obtained at the end of the MLDFT energy minimization are generally delocalized over the active and inactive regions. While this is not an issue for properties such as electronic excitation energies, for extensive response properties (e.g., polarizabilities and hyperpolarizabilities), such delocalization may lead to inaccurate results. It is therefore necessary to define occupied MOs that are spatially localized on the pre-defined fragment atoms. The resulting orbitals are referred to as Kohn-Sham Fragment-Localized MOs (KS-FLMOs).\cite{giovannini2024kohn,giovannini2025energy} They are obtained by minimizing the sum of the fragment energies ($E^A + E^B$) within the space spanned by the active and inactive occupied MOs. In this way, the total energy $E$ is kept constant, while the interaction term $E^{AB}_{\mathrm{int}}$, and thus the effective fragment–fragment repulsion, is maximized. \cite{giovannini2021energy,giovannini2022fragment,helgaker2014molecular} The resulting KS-FLMOs are thus maximally confined within the predefined $A$ and $B$ spatial regions. We remark that KS-FLMOs can also be defined in the context of full DFT calculations, and we refer the interested reader to Ref.~\citenum{giovannini2024kohn} for a detailed discussion of the computational procedure.

\begin{figure}
    \centering
    \includegraphics[width=.5\linewidth]{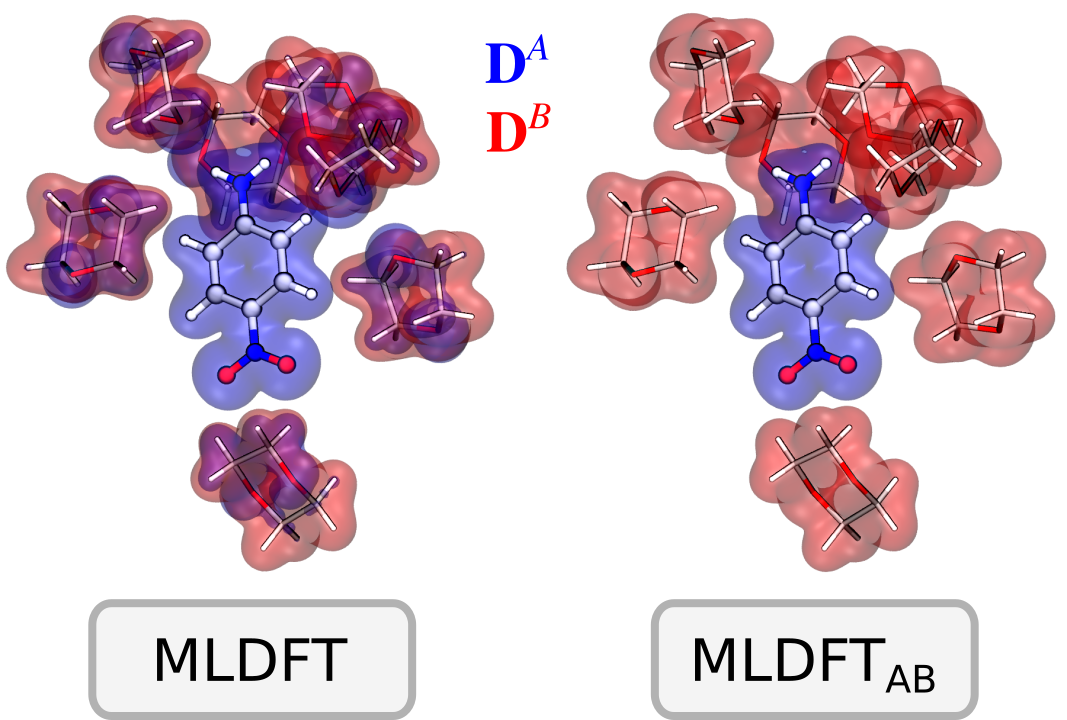}
    \caption{Active (A) and inactive (B) electron densities for a reduced snapshot of PNA in 1,4-dioxane as obtained within standard MLDFT using the partial Cholesky decomposition (left) and MLDFT$_{AB}$ based on KS-FLMOs (right). Isovalue: 0.02 a.u.}
    \label{fig:pna_dens}
\end{figure}

In the context of MLDFT, we refer to the resulting approach as MLDFT$_{AB}$, in line with our recent definition at the HF level.\cite{giovannini2021energy} To illustrate the localization achieved by the KS-FLMO procedure, Fig.~\ref{fig:pna_dens} compares the active and inactive densities obtained in standard MLDFT (i.e., using the partial Cholesky decomposition) and in MLDFT$_{AB}$ for a randomly selected reduced snapshot of PNA in 1,4-dioxane (7 solvent molecules). In both cases, PNA is the active region (CAM-B3LYP/aug-cc-pVDZ), while the solvent molecules represent the inactive subsystem (CAM-B3LYP/6-31G). In standard MLDFT, the active density partially extends into the inactive region, whereas in MLDFT$_{AB}$ it is properly confined within the active subsystem. We note that, since the total density matrix $\mathbf{D}$ is unchanged by the KS-FLMO procedure, MLDFT$_{AB}$ can be straightforwardly combined with a third (polarizable) MM layer without introducing additional embedding terms.

Before solving the response equations, the MLDFT energy is minimized again by starting from the fragment-localized active and inactive density matrices, to obtain active MOs that diagonalize the corresponding Fock matrix and can then be used for the subsequent evaluation of the response properties.\cite{giovannini2021energy}

\paragraph{MLDFT Response Theory}

In this work, we present for the first time an extension of MLDFT/FQ to the calculation of extensive molecular response properties, including linear polarizabilities and first hyperpolarizabilities. To this end, CPKS equations \cite{Casida95_155} must be reformulated for an MLDFT/FQ Hamiltonian. Building on the extension of MLDFT to time-dependent DFT (TD-DFT), \cite{giovannini2024tdmldft} we can write: \cite{Casida95_155,giovannini2024tdmldft}
\begin{equation}
 \label{eq:Casida}
\left[ \left (
 \begin{array}{cc}
  {\mathbf{A}} & {\mathbf{B}} \\
  {\mathbf{B}}^* & {\mathbf{A}}^* \\
 \end{array}
\right )
-
\omega
\left (
 \begin{array}{cc}
  -\bm{1}  & \bm{0} \\
  \bm{0}  & \bm{1} \\
 \end{array}
\right ) \right]
\left (
 \begin{array}{c}
  \mathbf{X} \\
  \mathbf{Y} \\
 \end{array}
\right ) = - \left (
 \begin{array}{c}
  \mathbf{Q} \\
  \mathbf{Q}^{*} \\
 \end{array}
\right )
\end{equation}
where $\omega$ is the frequency of the external electromagnetic field and $\mathbf{Q}$ is the property-specific right-hand side,\cite{lipparini2012linear} whose explicit form depends on the quantity of interest. For a generic hybrid functional, the $\mathbf{A}$ and $\mathbf{B}$ matrices are defined as:\cite{toulouse2013,gomez2023multiple}
\begin{align}
    {A}_{ai,bj} & = (\epsilon_a - \epsilon_i)\delta_{ab}\delta_{ij} + (ai|bj) - \cx (ab|ij) \nonumber \\ 
                & \quad - c_{\text{LR}} (ab|ij)_{\text{LR}} + f\ap{xc}_{ai,bj} + C^{\text{pol}}_{ai,bj} \label{eq:Amat} \\
    {B}_{ai,bj} & = (ai|bj) - \cx(aj|ib) \nonumber \\ 
                & \quad - c_{\text{LR}} (aj|ib)_{\text{LR}}  + f\ap{xc}_{ai,jb} + C^{\text{pol}}_{ai,jb} \label{eq:Bmat}
\end{align}
where $\epsilon$ denotes the MO energies (with $i,j$ labeling occupied MOs, $a,b$ virtual MOs, and $p,q,r,s$ general MOs), and $(pq|rs)$ are the two-electron integrals (the subscript LR indicating the range-separated form). We note that, within MLDFT, the MOs entering Eqs.~\ref{eq:Amat} and \ref{eq:Bmat} belong to the active region only, which results in a substantial reduction of the dimensionality of the response equations with respect to a full QM treatment. The term $C^{\text{pol}}$ collects the additional contributions to the $\mathbf{A}$ and $\mathbf{B}$ matrices arising from the polarizable embedding layer. For the specific case of QM/FQ, these contributions are given by:\cite{gomez2023multiple}
\begin{equation} \label{eq:Cpol_FQ}
    C\ap{FQ}_{ai,bj} = \sum_{p} \left(\int_{\mathbb{R}^3} \varphi_a(\mathbf{r}) \frac{1}{\abs{\mathbf{r}-\mathbf{r}_p}} \varphi_i(\mathbf{r}) \diff \mathbf{r} \right) \cdot \tilde{q}_{p} (\varphi_b,\varphi_j)
\end{equation}
where $\varphi$ denotes the molecular orbitals, and $\tilde{q}$ are the perturbed FQ charges located on MM atom $p$ at position $\mathbf{r}_p$, obtained by solving the following modified set of linear equations:\cite{lipparini2012linear} 
\begin{equation} \label{eq:linearqmfq_perturbed}
    \mathbf{M}\tilde{\mathbf{q}}_{\lambda} = - \mathbf{V}(\mathbf{X} + \mathbf{Y}) 
\end{equation}
which represents the perturbed counterpart of Eq.~\ref{eq:linearqmfq}. Within the MLDFT framework, the response associated with the frozen inactive density $\mathbf{D}^{B}$ is not included explicitly. This means that the response of the active subsystem is only indirectly affected by the inactive fragment through the relaxation of the active MOs arising from the ground-state solution. This may lead to an unphysical situation in which the polarizable FQ layer dynamically adjusts to the response of the active QM region, while the intermediate (inactive) MLDFT layer remains frozen.\cite{giovannini2024tdmldft,egidi2021polarizable} This is especially relevant for systems where environmental response contributions are significant. To properly account for the polarization of the inactive MLDFT layer, we resort to the strategy recently developed by us for electronic excitations in three-layer quantum-embedding/classical polarizable schemes, such as TD-MLDFT/FQ\cite{giovannini2024tdmldft} and FDE/FQ.\cite{egidi2021polarizable} Specifically, the inactive atoms are endowed with FQs exclusively when solving CPKS equations, so that they contribute to the polarization response at the same level as the outer classical FQ layer, thereby ensuring that the entire environment can dynamically respond to the external perturbation. We refer to this combined approach as MLDFT$^{\mathrm{pol}}_{AB}$/FQ, which merges the main elements of quantum embedding and polarizable QM/MM by incorporating electrostatics, polarization, and quantum repulsion effects. Notably, Pauli repulsion is treated explicitly at the ground-state level and affects the response implicitly through the relaxation of the ground-state MOs.

Once Eq.~\ref{eq:Casida} is solved, response properties can be evaluated. For the frequency-dependent polarizability $\boldsymbol{\alpha}(-\omega,\omega)$, the response equations in Eq.~\ref{eq:Casida} provide the first-order density response to a monochromatic external electric field oscillating at frequency $\omega$. The dynamic polarizability is then computed in the frequency domain as:\cite{rice1990frequency}%
\begin{gather}
    \boldsymbol{\alpha} (-\omega;\omega) = - 2~\sum_{ai} \boldsymbol{\mu}_{ai} \left( \mathbf{X}_{ai} + \mathbf{Y}_{ai} \right) 
\end{gather}
with $\boldsymbol{\mu}_{ai}$ denoting the virtual–occupied block of the electronic dipole moment operator. The dynamic polarizability describes the field-induced change of the molecular dipole moment under an oscillating electric perturbation and requires only the first-order density response. According to the $2n+1$ rule, the first-order perturbed density is also sufficient to evaluate non-linear properties, in particular the first hyperpolarizability. \cite{rice1990frequency,rice1992calculation,vangisbergen1998} Consequently, starting from the solution of Eq.~\ref{eq:Casida}, the frequency-dependent first hyperpolarizability can also be computed as:
\begin{gather} \label{eq:hyperpols-gen}
    \begin{split}
        \beta_{\eta \xi \tau} ( \omega_{1} ; \omega_{2}, \omega_{3} ) & = 2~\tr [ \mathbf{P}_{\eta}(-\omega_{1}) \boldsymbol{\Sigma}_{\xi \tau} ( -\omega_{2}, \omega_{3} ) ] \\
        & \quad + 2~\tr [ \mathbf{P}_{\xi}( \omega_{2}) \boldsymbol{\Sigma}_{\eta \tau} ( \omega_{1}, -\omega_{3} ) ] \\
        & \quad + 2~\tr [ \mathbf{P}_{\tau}( \omega_{3}) \boldsymbol{\Sigma}_{\eta \xi} ( -\omega_{2}, \omega_{1} ) ] \\
        & \quad - \tr [ \mathbf{k}\ap{xc}_{\xi \tau} ( \omega_{2}, \omega_{3}) (\mathbf{X}_{\eta}(\omega_1) + \mathbf{Y}_{\eta}(\omega_1))] 
    \end{split}
\end{gather}
where the frequencies $\omega_1, \omega_2$ and $\omega_3$ define the observables involved in non-linear processes: second-harmonic generation (SHG) process $\boldsymbol{\beta}(- 2 \omega; \omega, \omega)$, Pockels and Kerr effect $\boldsymbol{\beta}(- \omega; 0, \omega)$, and optical rectification $\boldsymbol{\beta}(0; \omega, -\omega)$, respectively. \cite{rice1990frequency,rice1992calculation,Baerends95_9347} In Eq. \ref{eq:hyperpols-gen}, the matrix $\mathbf{P}_{\eta}(\omega)$ contains the occupied-occupied and virtual-virtual block along the components $\eta = x,y,z$ at frequency $\omega$ and can be written as: \cite{rice1990frequency}
\begin{equation} \label{eq:m-operator}
    \begin{split}
        \mathbf{P}_{\eta,pq}(\omega) & = \boldsymbol{\mu}_{\eta,pq} + \left[2 (pq|bk) - \cx(pb|qk) \right. \nonumber \\
        & \left. \quad - \beta (ab|ij)_{\text{LR}} + f\ap{xc}_{pq,kb} +  C^{FQ}_{pq,bk} \right] \mathbf{X}_{\eta,bk}(\omega) \\ 
        & + \left[2 (pq|bk) - \cx(pk|qb) - \beta (ab|ij)_{\text{LR}} \right. \\
        & \left. \quad + f\ap{xc}_{pq,kb} +  C^{FQ}_{pq,bk}\right] \mathbf{Y}_{\eta,bk}(\omega)
    \end{split}
\end{equation}
while the matrix $\boldsymbol{\Sigma}(-\omega, \omega)$ is the second-order density that can be recast from the first-order densities as: \cite{rice1990frequency}
\begin{gather} \label{eq:p-2der}
    \boldsymbol{\Sigma}_{\eta \xi} (-\omega, \omega) =
    \begin{pmatrix}
        \mathbf{Y}_{\eta}^* (\omega) \mathbf{X}_{\xi} (\omega) +  \mathbf{Y}_{\xi}^* (\omega) \mathbf{X}_{\eta} (\omega) & \mathbf{0} \\
        \mathbf{0} & -\mathbf{Y}_{\eta} (\omega) \mathbf{X}_{\xi}^{*} (\omega) - \mathbf{Y}_{\xi} (\omega) \mathbf{X}_{\eta}^* (\omega)  \\
    \end{pmatrix}
\end{gather}
and analogously for the remaining frequency combinations in Eq.~\ref{eq:hyperpols-gen}. The final term in Eq.~\ref{eq:hyperpols-gen} contains the third-order derivative of the exchange-correlation energy functional $\mathbf{k}\ap{xc}_{\xi \tau} ( \omega_{2}, \omega_{3})$ with respect to the energy density, and originates from the nonlinear dependence of $v\ped{xc}$ on the density, as described in Ref.~\citenum{vangisbergen1998}. Within the MLDFT$_{AB}$/FQ and MLDFT$^{\text{pol}}_{AB}$/FQ framework, the perturbed densities $\mathbf{X}$, $\mathbf{Y}$, and $\boldsymbol{\Sigma}$ are defined in the space of the active orbitals only, but explicitly account for the interaction with both the inactive layer and the MM region. Remarkably, the polarizable MM portion instead directly modifies Eq.~\ref{eq:m-operator}. As a consequence, the MLDFT$_{AB}$/FQ and MLDFT$^{\text{pol}}_{AB}$/FQ polarizabilities and hyperpolarizabilities properly incorporate the electrostatic, polarization, and Pauli repulsion effects through embedding. 

\section{Computational Details} \label{sec:comp_det}

\begin{figure}[ht]
\centering
\includegraphics[width=.5\textwidth]{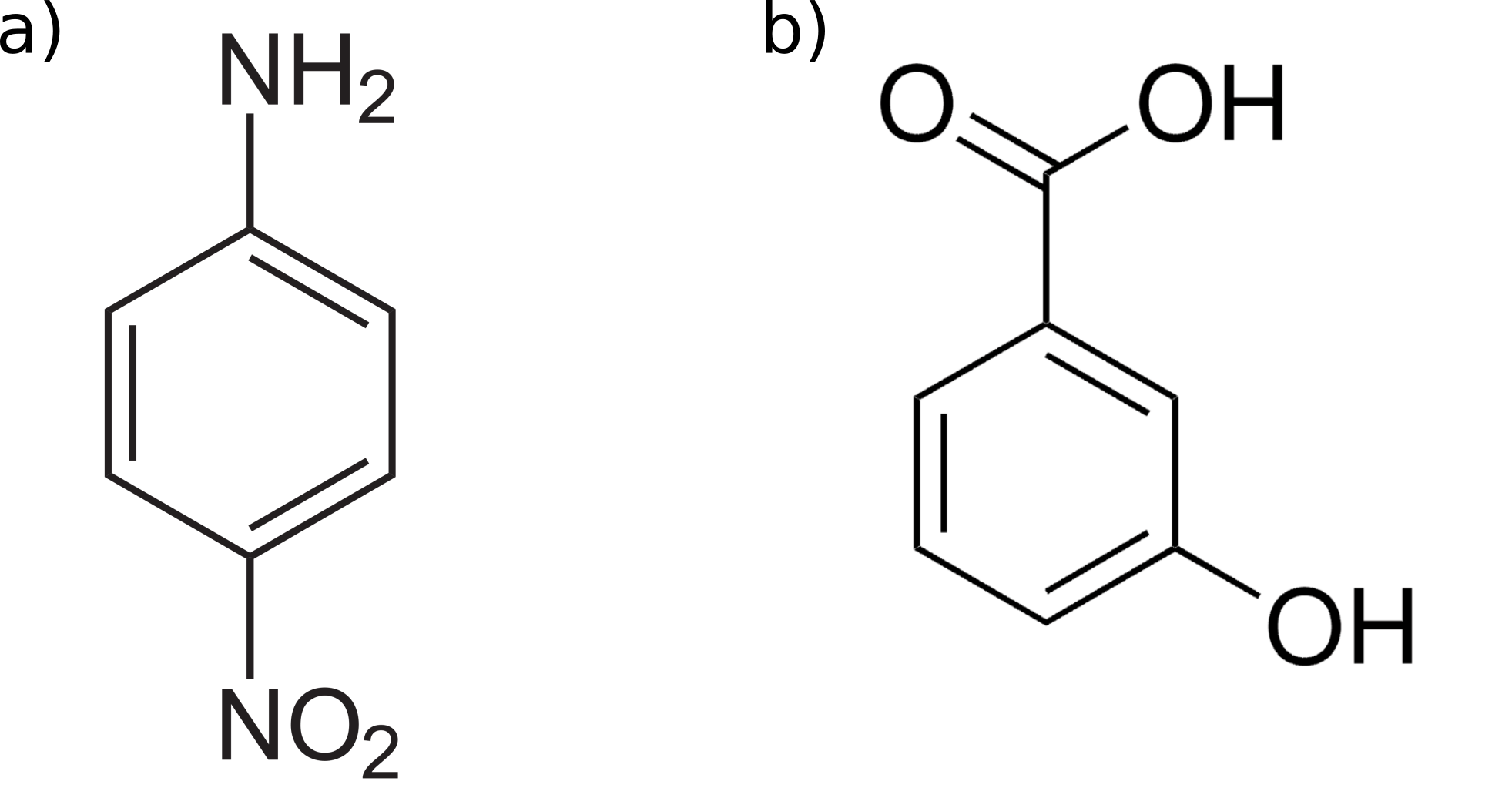}
\caption{Molecular structures of PNA (left) and HBA (right).}
\label{fig:molecules}
\end{figure}

The MLDFT$^{\mathrm{pol}}_{AB}$/FQ method is implemented in a development version of the electronic-structure program e$^{\mathcal{T}}$.\cite{eT2020,eT2025} As in Refs.~\citenum{marrazzini2021multilevel} and \citenum{giovannini2024tdmldft}, the DFT integration grid is built using the widely employed Lebedev angular quadrature\cite{lebedev1999quadrature} combined with the radial quadrature proposed in Ref.~\citenum{krack1998adaptive}. All calculations are performed using a 25th-order quadrature, with the radial threshold set to $10^{-5}$. The DFT functionals are implemented through an interface with the LibXC library.\cite{marques2012libxc,lehtola2018recent} In this work, MLDFT$^{\mathrm{pol}}_{AB}$/FQ is applied to two organic molecules, PNA and HBA, dissolved in 1,4-dioxane and in aqueous solution, respectively. PNA is chosen because it is a prototypical push–pull chromophore (see Fig. \ref{fig:molecules}a for its molecular structure) whose response to external environments has been extensively investigated both experimentally and theoretically.\cite{egidi2014benchmark,eriksen2013failures,frutos2013theoretical,kovalenko2000femtosecond,prabhumirashi1986solvent,millefiori1977electronic,champagne1996vibrational,jensen2005first,sok2011solvent,kosenkov2010solvent} PNA thus represents a perfect benchmark for assessing the ability of the newly developed embedding approaches to capture environmental effects on linear and non-linear response properties. As an additional test case, we consider HBA in aqueous solution, which features strong and heterogeneous solute–solvent interactions (see Fig. \ref{fig:molecules}b for its molecular structure).\cite{giovannini2018hyper,marrazzini2020calculation,ray1996comparative} For both molecules, experimental reference data are available,\cite{ray1996comparative,wortmann1993deviations,teng1983dispersion} allowing for an in-depth validation of the methods by direct comparison of the computed data with experiments.

Linear (polarizabilities) and non-linear (first hyperpolarizabilities) properties are computed by averaging the results performed on 21 uncorrelated frames extracted from the classical MD trajectories that are exploited to sample the solute-solvent phase space. This number is chosen as it guarantees in all cases that the properties report a maximum of about 3~\% statistical errors. The classical MD simulations are recovered from Refs. \citenum{ambrosetti2021quantum} (PNA) and \citenum{giovannini2018hyper} (HBA).
In both cases, the global hybrid B3LYP\cite{becke1988density,lee1988development,stephens1994ab} and range-separated CAM-B3LYP\cite{yanai2004new} functionals are exploited. The solute is treated as the active region and described by using the aug-cc-pVDZ (PNA) and 6-311++G$^{**}$ (HBA) basis sets, in agreement with previous studies,\cite{egidi2014stereoelectronic,marrazzini2020calculation,giovannini2018hyper} while the closest molecules define the inactive part in MLDFT and are described using the 6-31G basis set. 
The inactive region is chosen on the same criteria as our previous works,\cite{goletto2022linear,giovannini2024tdmldft} by including all solvent molecules within 2.5 \AA~(1,4-dioxane) and 3.5 \AA~(water) from each solute atom. All the remaining solvent molecules are described at the non-polarizable or FQ level, by exploiting the parameters reported in Ref. \citenum{ambrosetti2021quantum} (1,4-dioxane) and Ref. \citenum{giovannini2019eprdisrep} (water), respectively. 

\section{Numerical Applications} \label{sec:num_app}

In this section, we first discuss the dependence of linear and non-linear properties on the basis set exploited to describe the inactive region. We then present the numerical results for PNA solvated in 1,4-dioxane by using various embedding schemes, ranging from non-polarizable electrostatic embedding (EE) and polarizable QM/FQ, to MLDFT$_{AB}$/FQ and MLDFT$^{\text{pol}}_{AB}$/FQ. In particular, for the specific case of PNA in 1,4-dioxane, we first discuss the linear response by comparing static and dynamic isotropic polarizabilities, thereby assessing solvent-induced effects and dissecting the impact of polarization and inactive-layer contributions. We then move to the nonlinear response and analyze the different solvation approaches on the second-harmonic generation (SHG) first hyperpolarizability of both PNA in 1,4-dioxane and HBA in aqueous solution. In all cases, we discuss the accuracy of the various approaches by direct comparison with available experimental values.

\subsection{Dependence on the inactive region basis set}

\begin{figure}[!htbp]
    \centering
    \includegraphics[width=.5\linewidth]{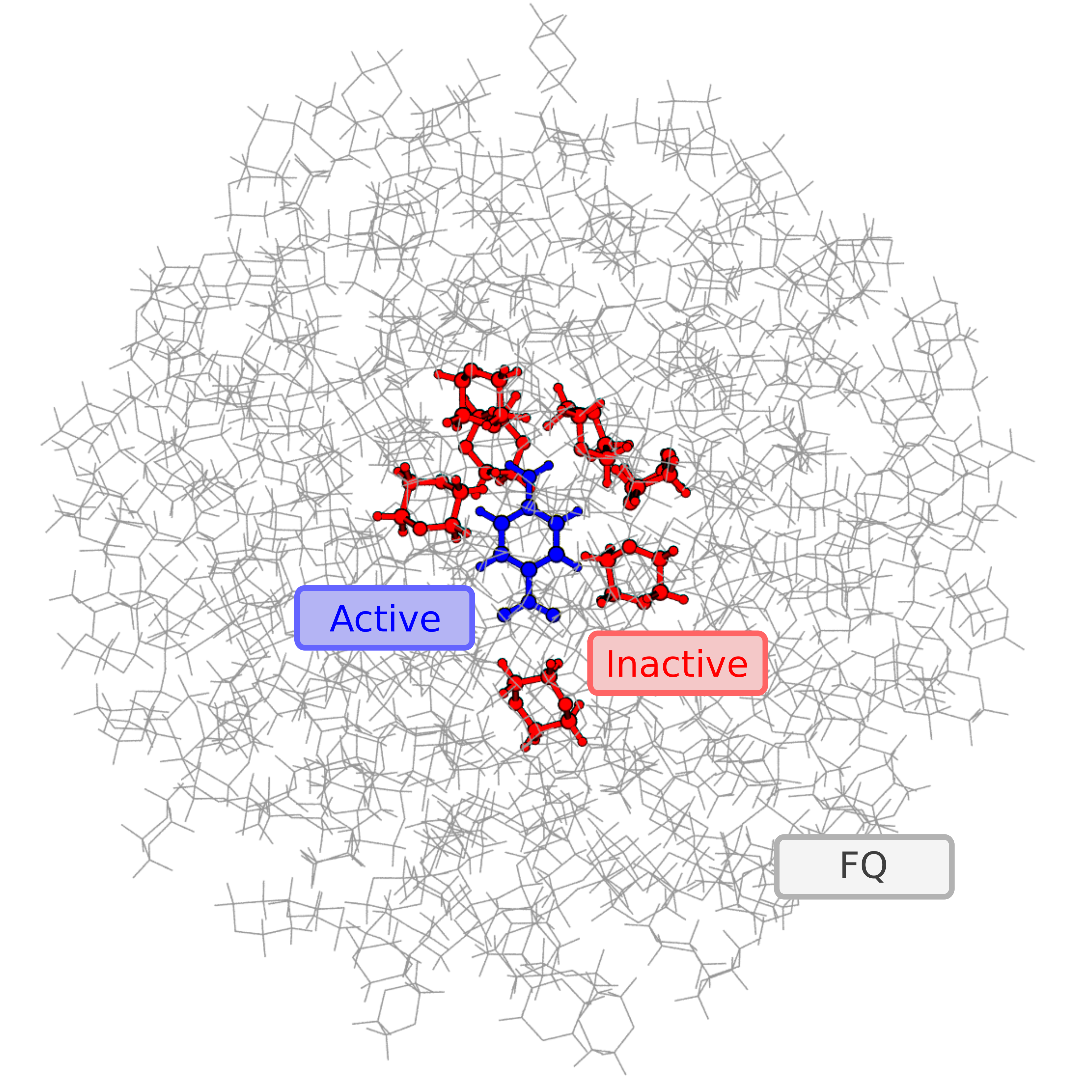}
    \caption{Graphical depiction of a randomly selected snapshot of PNA dissolved in 1,4-dioxane as partitioned at the MLDFT/FQ level.}
    \label{fig:PNA_structure}
\end{figure}

In MLDFT, the active and inactive regions can be described by using different basis sets. We then first evaluate the dependence of the computed linear and non-linear properties on the basis set used to describe the inactive region. To this end, we select a random snapshot of PNA in 1,4-dioxane, which is graphically depicted in Fig. \ref{fig:PNA_structure}. The inactive region is characterized by 7 solvent molecules, which are described by using 6-31G, 6-31G$^*$, and 6-31+G$^*$ basis sets, to quantify the inclusion of polarization and diffuse functions in the modeling of the inactive shell. In all cases, the MLDFT$^{\text{pol}}_{AB}$/FQ (CAM-B3LYP) level is exploited by treating the active region (PNA) using the aug-cc-pVDZ basis set.

\begin{table}[ht]
\centering
\small
\setlength{\tabcolsep}{4pt}
\caption{Calculated MLDFT$^{\text{pol}}_{AB}$/FQ (CAM-B3LYP) dipole moment (along the $z$-axis, Debye), isotropic polarizability (cm$^3$/mol), and static and dynamic ($\lambda = 1064$ nm) parallel components of the first hyperpolarizability with respect to the $z$-axis ($10^{-30}$ esu) of PNA in 1,4-dioxane, as a function of basis sets used for the inactive region. Relative errors (\%) with respect to the largest basis set (6-31+G$^*$) are given in parentheses.}
\label{tab:pna_dio_vs_basis}
\begin{tabular}{lrrr}
\hline
 & 6-31G & 6-31G* & 6-31+G* \\
\hline
    $\mu_z$ & $ 11.32~(2.79 \%)$ & $ 11.09~(0.75 \%)$ & $ 11.01$ \\
    \hline
    $\alpha(0;0)$  & $ 11.18~(0.65 \%)$ & $ 11.13~(0.94 \%)$ & $ 11.11$ \\
    $\beta_{\parallel}(0;0,0)$  & $ -19.46~(2.94 \%)$ & $ -19.08~(0.94 \%)$ & $ -18.90$ \\
    \hline
    $\alpha(-\omega;\omega)$ & $ 11.56~(0.74 \%)$ & $ 11.50~(0.22 \%)$ & $ 11.48$ \\
    $\beta_{\parallel}(-2\omega;\omega,\omega)$  & $ -37.42~(4.29 \%)$ & $ -36.37~(1.38 \%)$ & $ -35.88$ \\
\hline
\end{tabular}
\end{table}
In all calculations, PNA is oriented such that its main symmetry axis lies along the $z$-axis. To provide an overall comparison between the diverse basis-sets, we compute the $z$-component of the dipole moment, the isotropic polarizability, and the component of the first hyperpolarizability parallel to the $z$-axis as:
\begin{align}
    \alpha &= \frac{1}{3} \sum_{\xi = x,y,z} \alpha_{\xi \xi} \\
    \beta_{\parallel} &= \frac{1}{5} \sum_{\xi = x,y,z} \left(\beta_{\xi z\xi} + \beta_{z\xi\xi} + \beta_{\xi\xi z} \right) 
\end{align}
in the static regime ($\alpha(0;0)$, $\beta_\parallel(0;0,0)$) and the dynamic regimes ($\alpha(-\omega;\omega)$, $\beta_\parallel(-2\omega;\omega,\omega)$; $\lambda = 1064$ nm). 
The computed values are reported in Tab. \ref{tab:pna_dio_vs_basis}. By first looking at the dipole moments, we note that by enlarging the inactive basis set, the numerical value decreases, with a maximum discrepancy reported by 6-31G basis set of about 0.31 Debye. By moving to the static properties, the relative errors obtained with 6-31G with respect to 6-31+G$^*$ remain below $3~\%$, while in the dynamic regime they are below $5~\%$ for all quantities. Upon adding diffuse functions, the errors decrease to below $1~\%$ and $2~\%$ for 6-31G$^*$ in the static and dynamic regimes, respectively. Interestingly, both in the static and in the dynamic regimes, the hyperpolarizability is more sensitive to the inclusion of diffuse and polarization functions. However, the gain in accuracy is only marginal, and remarkably, it comes at a rapidly increasing computational demand. In particular, by moving from 6-31G to 6-31+G$^*$, the computational time increases by a factor of about 5. Accordingly, in the following, we use the 6-31G basis set for the inactive region, as it provides an overall good compromise between the computational cost and accuracy.

\subsection{para-Nitroaniline in 1,4-Dioxane}

We now move to discuss the numerical results obtained for PNA dissolved in 1,4-dioxane, which are computed by averaging the properties over 21 uncorrelated snapshots. We first focus on the static and dynamic linear response properties, such as electric dipole polarizabilities $\bm{\alpha}$. In particular, we compare the results obtained by modeling the system at the non-polarizable QM/MM (Electrostatic Embedding, QM/EE), polarizable QM/FQ, MLDFT$_{AB}$/FQ, and the polarizable MLDFT$^{\text{pol}}_{AB}$/FQ. Such a comparison allows us to evaluate and dissect the physicochemical mechanisms of the solute-solvent interactions captured by each solvation model, analyzing the relevance of electrostatics, polarization, and Pauli repulsion effects. For all embedding approaches, the response properties are evaluated using the global hybrid B3LYP and range-separated hybrid CAM-B3LYP functionals.

\begin{figure}[!htbp]
    \centering
    \includegraphics[width=.5\linewidth]{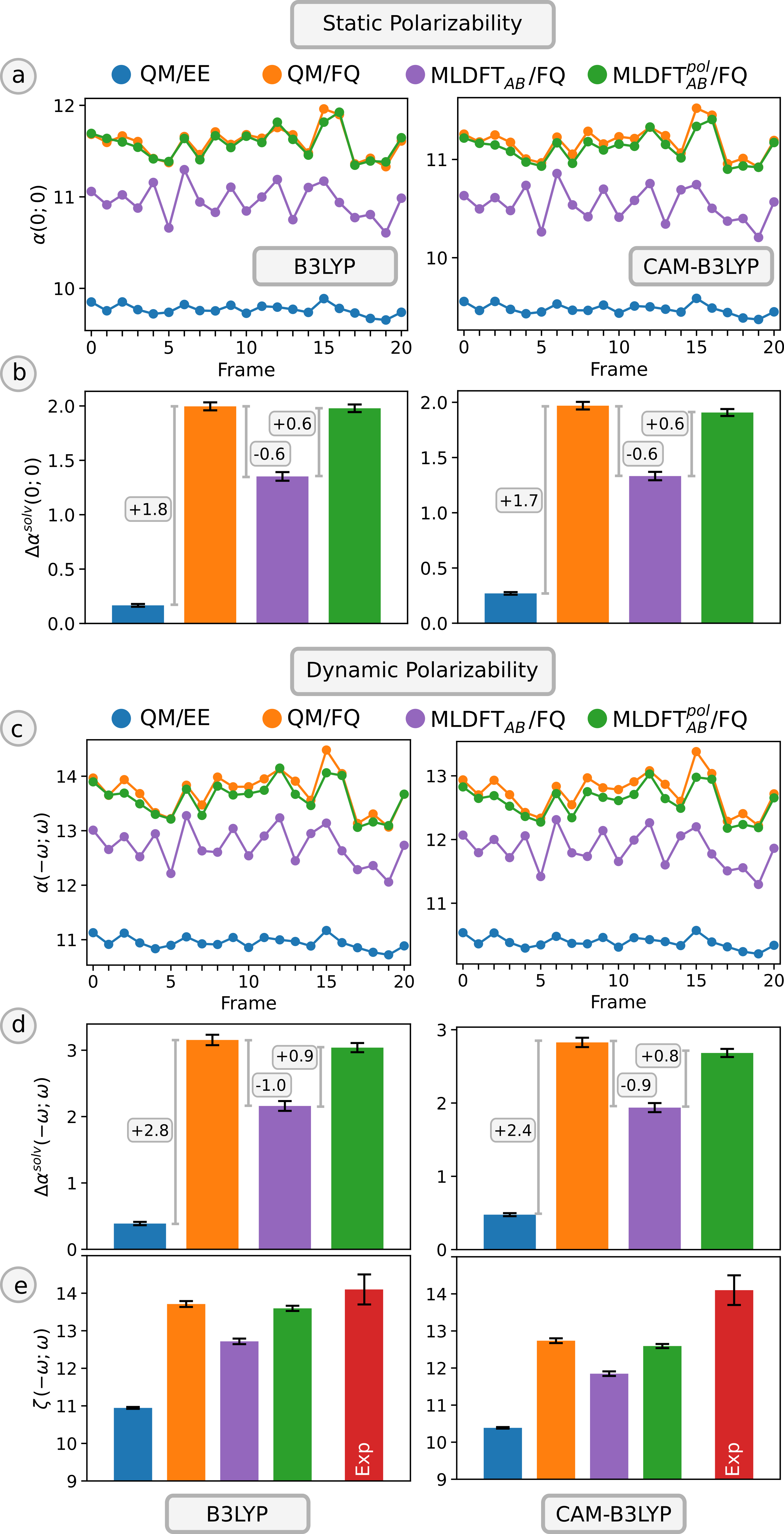}
    \caption{B3LYP (left) and CAM-B3LYP (right) isotropic polarizabilities of PNA in 1,4-dioxane, obtained by using the selected embedding methods. (a,c) Computed static (a) and dynamic (c) isotropic $\alpha$ as a function of the snapshot. (b,d) Computed solvent effects on static (b) and dynamic (d) isotropic $\alpha$ with respect to gas-phase value. (e) Comparison between computed and experimental (from Ref. \citenum{wortmann1993deviations}) isotropic molar polarizabilities $\zeta$. Error bars denote the statistical error. Dynamic values are obtained at $\lambda=589$ nm. All data are given in cm$^3$/mol.}
    \label{fig:PNA_polar_results}
\end{figure}

By first focusing on the static polarizabilities, in Fig. \ref{fig:PNA_polar_results}, we report the frame-by-frame values of the static isotropic polarizabilities $\alpha(0;0)$ obtained by the different embedding schemes and functionals. We first note a variability of all reported quantities with respect to the snapshots,\cite{goletto2022linear} highlighting the dependence of the properties on the specific environment fluctuations. Such fluctuations in the computed values are enhanced by the embedding methods that account for solute-solvent polarization, while at the QM/EE level, the variability is modest. This behavior reflects the higher sensitivity of the molecular response to the inclusion of the polarization effects along the trajectory. Remarkably, the relative ordering of the methods is almost constant within the various approaches: QM/MM systematically yields the smallest $\alpha$ values, while the inclusion of polarization and/or quantum embedding enhances the response. The inclusion of the inactive quantum embedding layer in MLDFT$_{AB}$/FQ provides an overall decrease of the computed values with respect to the purely electrostatic/polarizable QM/FQ. This highlights the role of quantum confinement introduced at the MLDFT level. However, it is interesting to note that the inclusion of mutual polarization effects in the response equation by means of MLDFT$^{\text{pol}}_{AB}$/FQ substantially increases MLDFT$_{AB}$/FQ towards the QM/FQ values. Indeed, MLDFT$_{AB}$/FQ yields slightly smaller $\alpha$ values than QM/FQ because of the inclusion of Pauli repulsion energy terms. Remarkably, these findings are coherently reproduced by both B3LYP and CAM-B3LYP functionals.

These trends are confirmed when averaging the raw data. In particular, we can appreciate the specificities of each embedding approach by computing the environment contribution to the isotropic polarizability $\Delta \alpha^{\text{solv}}$ as:
\begin{equation}
    \Delta \alpha^{\text{solv}} = \alpha^{\text{solv}} - \alpha^{\text{vac}}
\end{equation}
where $\alpha^{\text{solv}}$ is the isotropic polarizability averaged over the selected snapshots and $\alpha^{\text{vac}}$ is the isotropic polarizability of the PNA in vacuo (see also Tab. \ref{tab:PNA_alpha_0}). Numerical values for static $\Delta \alpha^{\text{solv}}$ are graphically depicted in Fig. \ref{fig:PNA_polar_results}b, together with their statistical errors. Solvent effects predicted by non-polarizable QM/MM are small ($\sim$ 0.2 cm$^{3}$/mol), while it becomes significantly larger when polarization and/or quantum embedding are employed. In particular, QM/FQ yields an increase of about 1.7-1.8 cm$^{3}$/mol at both B3LYP and CAM-B3LYP levels with respect to QM/EE. The introduction of the inactive layer in MLDFT$_{AB}$/FQ systematically reduces this contribution by approximately 0.6 cm$^{3}$/mol for both functionals, indicating a consistent decrease due to the inclusion of the Pauli repulsion. Moreover, with the inclusion of the inactive layer contribution in the active fragment response within the MLDFT$_{AB}^{\text{pol}}$/FQ scheme, a similar increase of about 0.6 cm$^{3}$/mol is observed for both functionals. The obtained results are consistent with the physicochemical mechanisms described by the various approaches.\cite{giovannini2019quantum} The fact that QM/FQ and MLDFT$_{AB}^{\text{pol}}$/FQ results almost coincide, highlights the mechanism of error cancellation in the former, which does not account for physical mechanisms such as Pauli repulsion. 

\begin{table}[!htbp]
    \centering
    \resizebox{\textwidth}{!}{%
    \begin{tabular}{cc|cccc|c}
        \hline
        \multicolumn{2}{c}{Method} & $\mu_z$ & $\alpha (0;0)$ & $\alpha^{\mu}$ & $\zeta (0;0)$ & $\zeta^{\text{exp}} (0;0)$ \\
        \hline
        \multirow{5}{*}{\rotatebox{90}{B3LYP}} & Gas-phase             & 7.70 & 9.60 & 287.22 & 296.82 & \multirow{10}{*}{404 $\pm$ 6} \\
                               & QM/EE                 & 9.76 $\pm$ 0.07   & 9.77 $\pm$ 0.01   & 462.35 $\pm$ 6.47  & 472.12 $\pm$ 6.48 & \\
                               & QM/FQ                 & 11.43 $\pm$ 0.13  & 11.60 $\pm$ 0.04  & 634.44 $\pm$ 13.87 & 646.04 $\pm$ 13.89 &\\
                               & MLDFT$_{AB}$/FQ       & 11.22  $\pm$ 0.11 & 10.95  $\pm$ 0.04 & 610.73 $\pm$ 11.88 & 621.69 $\pm$ 11.90 &\\
                               & MLDFT$^{\text{pol}}_{AB}$/FQ & 11.22  $\pm$ 0.11 & 11.58  $\pm$ 0.04 & 610.73 $\pm$ 11.88 & 622.31 $\pm$ 11.91 & \\
        \cline{1-6}
        \multirow{5}{*}{\rotatebox{90}{CAM-B3LYP}} & Gas-phase             & 7.40 & 9.21 & 264.57 & 273.78 &  \\
                                   & QM/EE                 & 9.48   $\pm$ 0.06 & 9.48  $\pm$ 0.01 & 435.75 $\pm$ 5.94  & 445.22 $\pm$ 5.95  & \\
                                   & QM/FQ                 & 10.99  $\pm$ 0.12 & 11.18 $\pm$ 0.03 & 587.17 $\pm$ 12.50 & 598.34 $\pm$ 12.53 & \\
                                   & MLDFT$_{AB}$/FQ       & 10.76  $\pm$ 0.10 & 10.54 $\pm$ 0.04 & 561.65 $\pm$ 10.70 & 572.18 $\pm$ 10.71 & \\
                                   & MLDFT$^{\text{pol}}_{AB}$/FQ & 10.76  $\pm$ 0.10 & 11.11 $\pm$ 0.03 & 561.65 $\pm$ 10.70 & 572.76 $\pm$ 10.72 & \\   
        \hline                                
    \end{tabular}
    }
    \caption{Calculated PNA in 1,4-dioxane $z$-component of the dipole moments ($\mu_z$ in Debye), isotropic static polarizabilities ($\alpha$ in cm$^{3}$/mol), reorientation terms ($\alpha^{\mu}$ in cm$^{3}$/mol), and total static molar polarizabilities ($\zeta$ in cm$^{3}$/mol). The experimental value (from Ref. \citenum{wortmann1993deviations}) is also provided, together with gas-phase results.}
    \label{tab:PNA_alpha_0}
\end{table}

To compare our results with the available experiments, we compute the total isotropic static molar polarizabilities as:\cite{wortmann1993deviations,egidi2014stereoelectronic,goletto2022linear}
\begin{equation} \label{eq:static_alpha}
	\zeta(0;0) = \alpha (0;0)+ \alpha^{\mu}
\end{equation}
where $\alpha^\mu$ represents the isotropic reorientation term, which reads:
\begin{equation}
	\alpha^{\mu} = \frac{|\bm{\mu}|^2}{3k_{\mathrm{B}}T}
\end{equation}
where $\bm{\mu}$ is the dipole moment, $k_\mathrm{B}$ is the Boltzmann constant, and $T$ is the temperature. In Tab. \ref{tab:PNA_alpha_0}, the $z$-component of the dipole moment, the isotropic static polarizabilities, the reorientation terms, and total static molar polarizabilities obtained by using the various embedding approaches are reported together with their computed gas-phase and experimental counterparts from Ref. \citenum{wortmann1993deviations}. 
Moving from the gas phase to solution, all embedding schemes predict an increase of $\mu_z$ and $\alpha(0;0)$, highlighting the effect of the environment on the ground state and response properties. Such an increase is generally small in the case of the QM/EE embedding, while the largest values are obtained for the QM/FQ model. The inclusion of the inactive layer through the MLDFT scheme has only a minor impact on $\mu_z$. We remark that the two MLDFT variants only differ by the inclusion of polarization effects in the response equations, and thus they predict the same $\mu_z$. The reorientation term $\alpha^{\mu}$ dominates the total response and largely defines the total static molar polarizabilities $\zeta(0;0)$. Indeed, this quantity strongly depends on the computed $\mu_{z}$ (in all frames, the main molecular axis is collinear with the z axis). By comparing the total static molar polarizabilities $\zeta(0;0)$ with the experimental reference value, the closest values are obtained for the QM/EE embedding, while all other schemes yield numerical results that substantially overestimate the experiment. This is particularly evident for B3LYP, which predicts $\zeta(0;0)$ values that are larger than CAM-B3LYP in all cases. The reported discrepancy with respect to the experiment is primarily due to the differences in the computed dipole moments since $\alpha(0;0)$ only slightly affects the final computed property. Therefore, the most important error source for the polarizable/quantum embedding approaches lies in the overestimation of $\mu_z$, which is enhanced when the polarization is included in the modeling, and lowered by the Pauli repulsion effects introduced in multilevel methods. Such a trend is in perfect agreement with Ref. \citenum{goletto2022linear}, where some of us computed the same property using highly correlated methods, such as Coupled-Cluster singles and (perturbative) doubles (CC2 and CCSD), embedded in a HF wavefunction for the environment.

To eliminate the dependence of the results on the computed dipole moments, we can move to the dynamic polarizabilities ($\lambda$ = 589 nm), for which the reorientation term $\alpha^\mu = 0$.\cite{wortmann1993deviations,cammi2000attempt,egidi2014stereoelectronic} The dependence of the dynamic property as a function of the snapshot is reported in Fig. \ref{fig:PNA_polar_results}c, while the associated $\Delta\alpha^{\text{solv}}$ are given in Fig. \ref{fig:PNA_polar_results}d, at both the B3LYP and CAM-B3LYP levels. All the trends perfectly align with those reported for the static polarizability: (i) polarizable approaches display the largest variability along the trajectory approaches; (ii) the inclusion of mutual solute-solvent polarization (QM/EE $\rightarrow$ QM/FQ) increases the computed polarizability by about 2.8 (B3LYP) and 2.4 (CAM-B3LYP) cm$^{3}$/mol; (iii) Pauli repulsion effects (QM/FQ $\rightarrow$ MLDFT$_{AB}$/FQ) have an opposite effect, decreasing the computed property by about 1.0 cm$^{3}$/mol; (iv) including polarization effects in the response equations has a similar, but opposite effect to accounting for Pauli repulsion effects.

The total dynamic isotropic molar polarizabilities $\zeta(-\omega;\omega) = \alpha(-\omega;\omega)$ for the different embedding schemes are graphically given in Fig. \ref{fig:PNA_polar_results}e, where their experimental counterpart\cite{wortmann1993deviations} is also given. The comparison with experiment shows that QM/EE yields the largest discrepancy, whereas QM/FQ provides the closest agreement for both functionals. Notably, incorporating inactive-layer response effects within MLDFT$_{AB}^{\text{pol}}$/FQ leads to values that are in better agreement with experiment than those obtained with MLDFT$_{AB}$/FQ, and is comparable to QM/FQ, which provides a good agreement also thanks to error cancellation. In this context, B3LYP yields $\alpha(-\omega;\omega)$ values consistent with the experimental reference, whereas CAM-B3LYP systematically underestimates the measured polarizability. 

We now move to the calculation of the SHG first hyperpolarizability for PNA dissolved in 1,4-dioxane evaluated at a wavelength $\lambda = 1064$ nm, which has been experimentally measured in Ref. \citenum{teng1983dispersion} by means of Electric Field Induced SHG (EFISHG). To compare the computed data with the experimental result, we calculate:\cite{willetts1992problems} 
\begin{equation}
    \beta^B_z(-2\omega;\omega,\omega) = \dfrac{1}{3} \sum_{\xi=x,y,z} (\beta_{z\xi\xi} + \beta_{\xi z\xi} + \beta_{\xi \xi z})
\end{equation}
where we use the perturbation series convention discussed in Ref. \citenum{willetts1992problems} to allow for a direct comparison with EFISHG measured data.

\begin{figure}[!htbp]
    \centering
    \includegraphics[width=.5\linewidth]{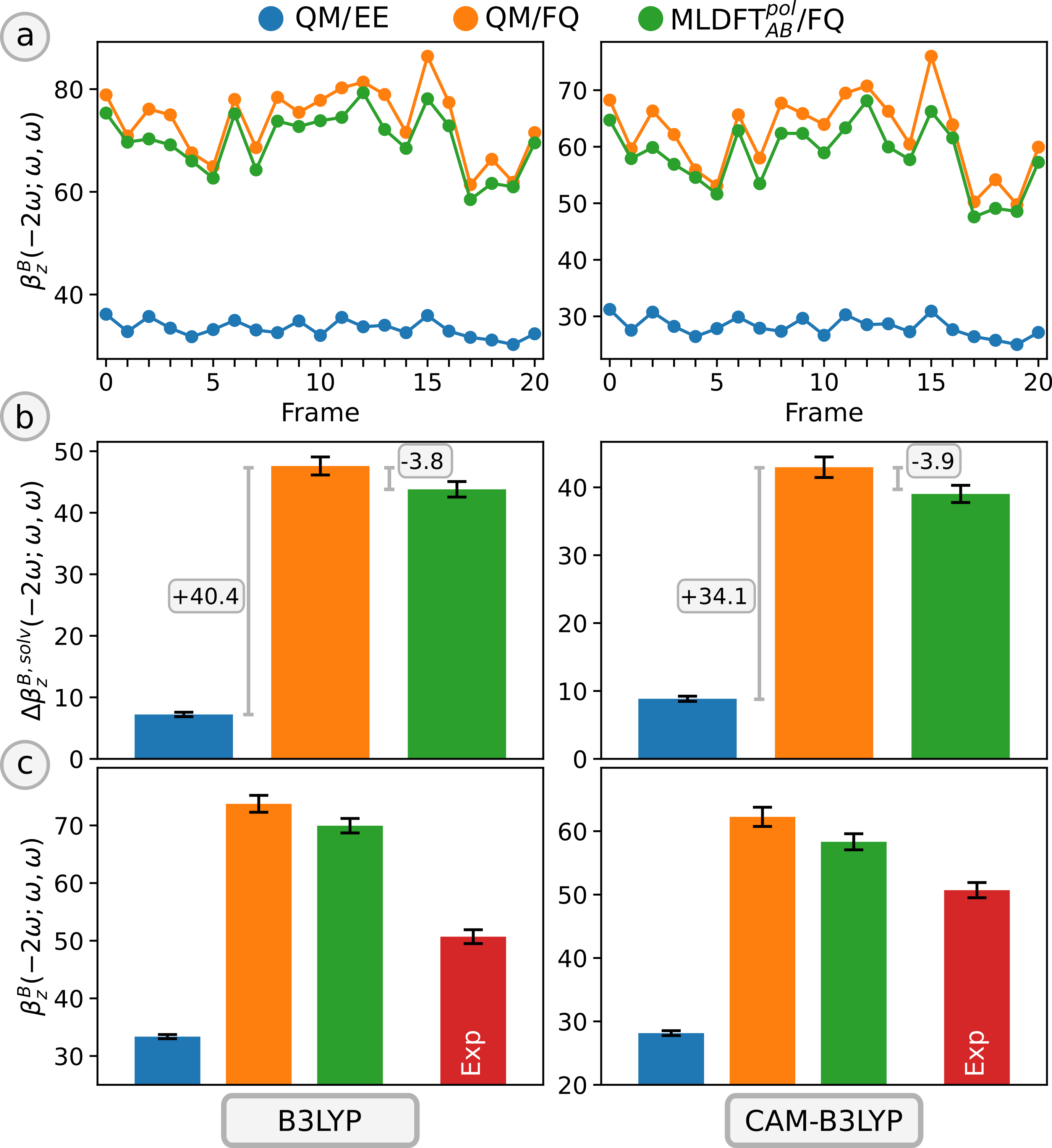}
    \caption{B3LYP (left) and CAM-B3LYP (right) SHG $\beta^B_z$ values of PNA in 1,4-dioxane obtained by using the selected embedding methods. (a) Computed $\beta^B_z$ as a function of the snapshot. (b) Computed solvent effects on $\beta^B_z$ with respect to gas-phase value. (c) Comparison between computed and experimental (from Ref. \citenum{teng1983dispersion}) $\beta^B_z$. Error bars denote the statistical error. All data are computed at $\lambda = 1064$ nm and are given in esu.}
    \label{fig:PNA_hyperpolar_results}
\end{figure}

Fig. \ref{fig:PNA_hyperpolar_results}a reports the $\beta^B_z(-2\omega;\omega,\omega)$ values obtained by using B3LYP (left panels) and CAM-B3LYP (right panels) functionals as a function of the snapshots. In this case, we consider non-polarizable QM/EE, polarizable QM/FQ, and MLDFT$^{\text{pol}}_{AB}$/FQ. As observed for polarizability, the frame-by-frame values show a large dependence on the configuration of the solvent, especially when polarizable and quantum embedding are employed. In fact, QM/FQ and MLDFT$^{\text{pol}}_{AB}$/FQ fluctuate significantly along the trajectory between 60 and 85 esu for B3LYP and between 50 and 70 esu for CAM-B3LYP. In contrast, QM/EE yields overall much smaller values and a narrower dispersion, indicating that a purely electrostatic embedding captures only a limited portion of the solvent-induced modulation of the nonlinear response. The relative ordering is in line with the results obtained for polarizability. Again, MLDFT$^{\text{pol}}_{AB}$/FQ remains slightly below QM/FQ along the frames, consistent with the inclusion of Pauli repulsion effects introduced by the inactive-layer.

In Fig. \ref{fig:PNA_hyperpolar_results}b we graphically depict the solvent effect on the computed $\beta_{z}^{B} (-2 \omega; \omega, \omega)$, which is calculated as:
\begin{equation}
    \Delta \beta_{z}^{B} (-2 \omega; \omega, \omega) ={\beta}_{z}^{B,\text{solv}} (-2 \omega; \omega, \omega) -\beta_{z}^{B,\text{vac}} (-2 \omega; \omega, \omega)
    \label{eq:delta_beta_solv}
\end{equation}
where $\beta_{z}^{B,\text{solv}}$ and $\beta_{z}^{B,\text{vac}}$ are the SHG hyperpolarizability computed in solution (averaged over the snapshots) and in the gas-phase. Raw data are given in Tab. S2 in the SI. While QM/EE predicts only a modest solvent-induced increase with respect to the gas phase, QM/FQ yields a much larger enhancement. The gap between these two descriptions is approximately $+40.4$ esu for the B3LYP functional and $+34.1$ esu for the CAM-B3LYP functional, highlighting the critical role of mutual polarization between solute and environment. This represents an increase of approximately 550~\% (B3LYP) and 400~\% (CAM-B3LYP). By including the Pauli repulsion effect by means of MLDFT$^{\text{pol}}_{AB}$/FQ, a reduction of about $3.8$--$3.9$ esu with respect to purely classical QM/FQ is reported, highlighting the competing role of quantum confinement versus polarization and electrostatic effects. Although with a lower extent with respect to polarization, as expected, the inclusion of Pauli repulsion yields a non-negligible decrease of about 8~\% and 10~\% for B3LYP and CAM-B3LYP functionals, respectively.
Finally, Fig. \ref{fig:PNA_hyperpolar_results}c compares the averaged $\beta_{z}^{B} (-2 \omega; \omega, \omega)$ with the experimental reference value from Ref. \citenum{teng1983dispersion}. For both functionals, electrostatic QM/EE largely underestimates the experimental counterpart. Including mutual polarization effects by means of QM/FQ shifts the computed hyperpolarizability towards experiment. However, independently of the DFT functional exploited, the computed hyperpolarizability overestimates its experimental counterpart by about 45~\% (B3LYP) and 23~\% (CAM-B3LYP). Quantum repulsion effects by means of MLDFT$^{\text{pol}}_{AB}$/FQ yield in both cases the best agreement with experiment, reducing the computing error to 38~\% and 15~\% for B3LYP and CAM-B3LYP, respectively. Specifically, as expected, CAM-B3LYP is generally closer to the experimental counterpart, highlighting the role of a proper treatment of long-range electron interactions in the prediction of non-linear properties of push-pull chromophores, such as PNA. 

\subsection{3-Hydroxybenzoic Acid in Aqueous Solution}

We now move to study HBA dissolved in aqueous solution, for which the second-harmonic hyperpolarizability has been determined by hyper-Rayleigh scattering (HRS)\cite{clays1991hyper,clays1992hyper} experiments ($\lambda = 1064$ nm).\cite{ray1996comparative} In agreement with previous studies based on the same experiment,\cite{ray1996comparative,giovannini2018hyper,marrazzini2020calculation} we compare our calculated results with the experiments by computing:

\begin{equation}
    \beta_{\text{HRS}}(-2\omega;\omega,\omega) = \sqrt{\sum_{\xi=x,y,z} \left(\sum_{\eta=x,y,z} \beta_{\xi \eta\eta} + \beta_{\eta \xi \eta} + \beta_{\eta\eta \xi} \right)}
    \label{eq:shg_hrs}
\end{equation}

\begin{figure}[!htbp]
    \centering
    \includegraphics[width=.5\linewidth]{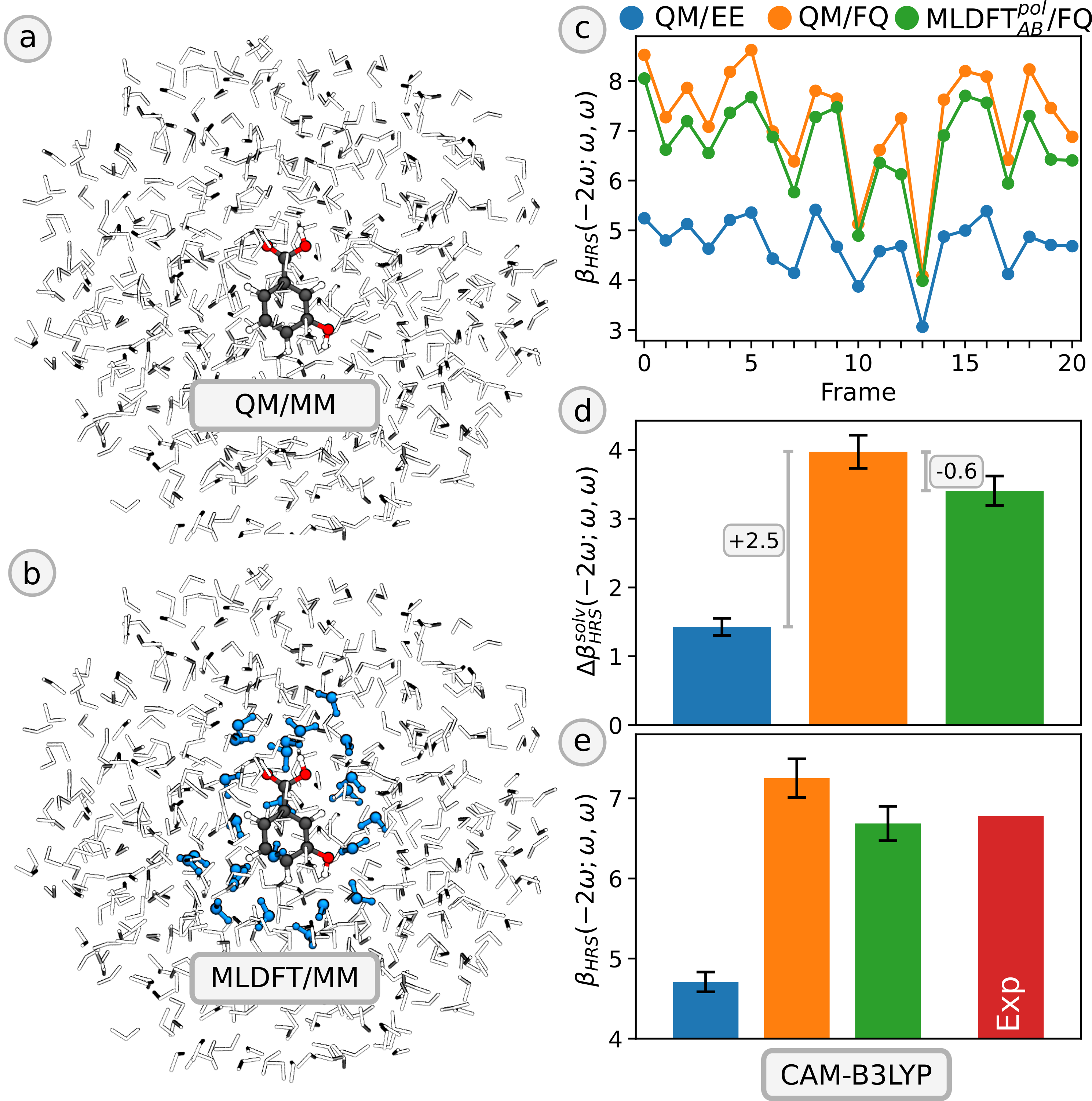}
    \caption{(a-b) Graphical depiction of HBA in aqueous solution as described by QM/MM  (a) and MLDFT/MM (b) methods. (c) Computed SHG $\beta_{\text{HRS}}$ as a function of the snapshot. (d) Computed solvent effects on SHG $\beta_{\text{HRS}}$ with respect to gas-phase value. (e) Comparison between computed and experimental (from Ref. \citenum{ray1996comparative}) SHG $\beta_{\text{HRS}}$. Error bars denote the statistical error. All data are computed using CAM-B3LYP functional at $\lambda = 1064$~nm and are given in esu.}
    \label{fig:HBA}
\end{figure}

As for the case of PNA in 1,4-dioxane, we consider various approaches for treating environmental effects on response properties, ranging from non-polarizable QM/EE, to polarizable QM/FQ, and MLDFT$^{\text{pol}}_{AB}$/FQ level, where the polarization of the inactive shell in the response equations is treated at the FQ level. In Fig. \ref{fig:HBA}a-b, we graphically depict a randomly selected snapshot of HBA dissolved in water as treated at the QM/MM level (a) and by using a three-layer MLDFT/MM approach (b), where the inactive water molecules are highlighted in blue. The computed results by exploiting all methods are graphically depicted in Fig. \ref{fig:HBA}c-e. In particular, in Fig. \ref{fig:HBA}c we graphically depict the raw data computed for each frame extracted from the classical MD trajectory. Similar to PNA, the values of the computed $\beta_{\text{HRS}}$ vary along the frames, ranging between 3 and 5.5 esu for QM/MM, and 4 and 8.5 esu for both QM/FQ and MLDFT$^{\text{pol}}_{AB}$/FQ. This again underscores the variability between solute-solvent interactions in the MD trajectory. Interestingly, we note a trend consistent with our findings on PNA: QM/MM values are always lower than the corresponding polarizable and quantum embedding counterparts. At the same time, MLDFT$^{\text{pol}}_{AB}$/FQ predicts $\beta_{\text{HRS}}$ values that are lower than QM/FQ frame by frame, highlighting a consistent contribution of Pauli repulsion effects introduced by the inclusion of the inactive layer in MLDFT. 

To quantify the solvent effects on the computed property, we calculate the solvent variation of the property $\Delta \beta^{\text{solv}}_{\text{HRS}}$ (see Eq. \ref{eq:delta_beta_solv}) by using the value in the gas-phase reported in Ref. \citenum{marrazzini2020calculation} as a reference (see Tab. S3 in the SI). The computed $\Delta \beta^{\text{solv}}_{\text{HRS}}$ values together with the associated statistical errors are reported in Fig. \ref{fig:HBA}d. We note that the inclusion of mutual polarization drastically increases the contribution of solvent effects by almost doubling that predicted at the non-polarizable QM/MM level. At the same time, confinement effects included at the MLDFT level decrease the QM/FQ value by about 0.6 esu, which corresponds to a non-negligible 15~\% decrease of the purely polarizable contribution. 

In Fig. \ref{fig:HBA}e, the comparison between computed and experimental data from Ref. \citenum{ray1996comparative} is graphically depicted. As it can be appreciated, the non-polarizable QM/EE approach systematically underestimates the experimental reference by about 30~\%. The inclusion of purely electrostatic polarization by means of QM/FQ moves the computed hyperpolarizability towards the experimental data, however, overestimating it by about 7~\%. Finally, MLDFT$^{\text{pol}}_{AB}$/FQ gives an almost perfect agreement with experiment, thanks to the additional decrease of the property given by electronic confinement. We remark that similar findings are also obtained by using the B3LYP functional to describe the QM (MLDFT) region. In this case, the absolute values are generally larger than those computed at the CAM-B3LYP level, as previously shown for PNA, and the agreement with the experimental result is less satisfactory. Still, the best performing method is MLDFT$^{\text{pol}}_{AB}$/FQ, which, thanks to a quantum-based description of Pauli repulsion, moves the computed results towards the experiment. 

\section{Summary and Conclusions} \label{sec:sum_conc}

In this work, we have introduced the extension of polarizable MLDFT/MM to compute extensive molecular properties in complex environments. The protocol is formulated in terms of CPKS equations, which are written for the polarizable MLDFT/FQ Hamiltonian, consistently incorporating the mutual polarization between the QM active fragment and the polarizable embedding. To enforce a physically meaningful localization of the active occupied orbitals, we combine the procedure with KS-FLMOs,\cite{giovannini2024kohn} which allow us to prevent spurious delocalization across the active–inactive partitioning. In addition, we have introduced an additional polarization term for the inactive MLDFT layer at the response level. In the resulting MLDFT$^{\mathrm{pol}}_{AB}$/FQ scheme, the entire environment (inactive layer + outer MM region) can dynamically respond to the external perturbation, while Pauli repulsion remains accounted for at the ground-state level and affects the response through the relaxation of the active MOs.

We have applied the proposed framework to two prototypical systems, PNA in 1,4-dioxane and HBA in water, considering both linear polarizabilities and first hyperpolarizabilities in the static and frequency-dependent regimes. The numerical analysis highlights that electrostatic embedding alone captures only a limited portion of solvent effects, whereas the inclusion of mutual polarization leads to substantial enhancements of both linear and non-linear responses. At the same time, the introduction of an inactive quantum layer through MLDFT produces an opposing contribution associated with quantum confinement effects, consistent with the role of Pauli repulsion in reshaping the electronic density of the active part. By explicitly including the inactive-layer polarization in the response equations (MLDFT$^{\mathrm{pol}}_{AB}$/FQ), we obtain a balanced description in which polarization and confinement effects are simultaneously represented, enabling a controlled dissection of the relative contributions of electrostatics, polarization, and quantum repulsion to the final observable. Overall, our findings demonstrate that the MLDFT$^{\mathrm{pol}}_{AB}$/FQ provides a good agreement with the available experimental reference. 

More generally, the present work provides a transferable route for computing response properties of embedded systems within a fully atomistic, multilevel quantum-embedding/polarizable MM framework. Importantly, the protocol is not limited to polarizabilities and first hyperpolarizabilities and can be systematically extended to other electric and mixed response properties, by appropriate definitions of the property-specific right-hand side and response functionals. In this perspective, MLDFT$^{\mathrm{pol}}_{AB}$/FQ offers a general platform for the quantitative modeling of response properties in complex environments, combining the efficiency of a reduced MO space with an explicit treatment of environmental electrostatics, polarization, and quantum confinement effects.

\begin{acknowledgement}
This work was supported by the European Research Council (ERC) under the European Union’s Horizon 2020 Research and Innovation Program (grant agreement No. 101020016). This work was funded by the European Union – Next Generation EU in the framework of the PRIN 2022 PNRR project POSEIDON – Code P2022J9C3R. This publication is also based upon work of COST Action CA21101 “Confined molecular systems: from a new generation of materials to the stars” (COSY) supported by COST (European Cooperation in Science and Technology). 
\end{acknowledgement}

\begin{suppinfo}
Raw data for PNA in 1,4-dioxane (dynamic polarizabilities and hyperpolarizabilities) and HBA in aqueous solution (dynamic polarizabilities) calculated using B3LYP and CAM-B3LYP functionals.
\end{suppinfo}

{
\small
\bibliography{bibliography}

\begin{mcitethebibliography}{119}
\bibitem[Lin and Truhlar(2007)Lin, and Truhlar]{lin2007qm}
Lin,~H.; Truhlar,~D.~G. {QM/MM}: what have we learned, where are we, and where do we go from here? \emph{Theor. Chem. Acc.} \textbf{2007}, \emph{117}, 185--199\relax
\mciteBstWouldAddEndPuncttrue
\mciteSetBstMidEndSepPunct{\mcitedefaultmidpunct}
{\mcitedefaultendpunct}{\mcitedefaultseppunct}\relax
\EndOfBibitem
\end{mcitethebibliography}
}

\end{document}